\documentclass[conference]{IEEEtran}

\usepackage{url}
\usepackage{cite}
\usepackage{graphicx} 
\usepackage{subcaption}
\usepackage{mwe}
\usepackage{pbox}
\usepackage{enumerate}
\usepackage{booktabs}
\usepackage{listings}
\usepackage{color}
\usepackage{amsmath}
\usepackage{amsfonts} 
\usepackage{float}
\usepackage{algpseudocode}
\usepackage{algorithm}
\usepackage{lipsum}
\usepackage{comment} 

\usepackage{tikz}
\usepackage{textcomp}
\usepackage{hyperref}
\usepackage{lipsum}

\newcommand\copyrighttext{%
  \footnotesize \textcopyright 2020 IEEE. Personal use of this material is permitted. Permission from IEEE must be obtained for all other uses, in any current or future media, including reprinting/republishing this material for advertising or promotional purposes, creating new collective works, for resale or redistribution to servers or lists, or reuse of any copyrighted component of this work in other works.
  DOI: \href{https://doi.org/10.1109/SCAM51674.2020.00022}{10.1109/SCAM51674.2020.00022}}
\newcommand\copyrightnotice{%
\begin{tikzpicture}[remember picture,overlay]
\node[anchor=south,yshift=10pt] at (current page.south) {\fbox{\parbox{\dimexpr\textwidth-\fboxsep-\fboxrule\relax}{\copyrighttext}}};
\end{tikzpicture}%
}

\algnewcommand\algorithmicforeach{\textbf{for each}}
\algdef{S}[FOR]{ForEach}[1]{\algorithmicforeach\ #1\ \algorithmicdo}

\lstnewenvironment{code}[1][]%
  {\noindent\minipage{\linewidth}\medskip 
   \lstset{basicstyle=\ttfamily\footnotesize,frame=single,#1}}
  {\endminipage}

\definecolor{dkgreen}{rgb}{0,0.6,0}
\definecolor{gray}{rgb}{0.5,0.5,0.5}
\definecolor{mauve}{rgb}{0.58,0,0.82}

\title{\LARGE \bf
DepGraph: Localizing Performance Bottlenecks in Multi-Core Applications Using Waiting Dependency Graphs and Software Tracing
}

\author{\IEEEauthorblockN{Naser Ezzati-Jivan}
\IEEEauthorblockA{
Brock University\\
nezzatijivan@brocku.ca}
\and
\IEEEauthorblockN{Quentin Fournier}
\IEEEauthorblockA{Polytechnique Montreal\\
quentin.fournier@polymtl.ca}
\and
\IEEEauthorblockN{Michel R. Dagenais}
\IEEEauthorblockA{Polytechnique Montreal\\
michel.dagenais@polymtl.ca}
\and
\IEEEauthorblockN{Abdelwahab Hamou-Lhadj}
\IEEEauthorblockA{Concordia University\\
wahab.hamou-lhadj@concordia.ca
}}

\begin{document}

\maketitle
\thispagestyle{empty}
\pagestyle{empty}

%
\copyrightnotice

\begin{abstract}
This paper addresses the challenge of understanding the waiting dependencies between the threads and hardware resources required to complete a task. The objective is to improve software performance by detecting the underlying bottlenecks caused by system-level blocking dependencies. In this paper, we use a system level tracing approach to extract a Waiting Dependency Graph that shows the breakdown of a task execution among all the interleaving threads and resources. The method allows developers and system administrators to quickly discover how the total execution time is divided among its interacting threads and resources. Ultimately, the method helps detecting bottlenecks and highlighting their possible causes. Our experiments show the effectiveness of the proposed approach in several industry-level use cases. Three performance anomalies are analysed and explained using the proposed approach. Evaluating the method efficiency reveals that the imposed overhead never exceeds 10.1\%, therefore making it suitable for in-production environments.
\end{abstract} 

\section{Introduction}

Analysing performance bottlenecks and identifying their root causes in multi-core systems is known to be a challenging task~\cite{introperf}. Performance anomalies may arise due to a wide range of reasons such as a bug in the code, a flawed database design, a misconfiguration, a lack of resources, or the overall system load. In a multi-core system, however, performance issues are mostly due to the number of concurrent threads and their complex interactions, combined with the need to share common resources which also often leads to resource contention problems ~\cite{introperf}. 

Throughout the development phase, software developers have many tools (such as interactive debuggers) at their disposal to detect and analyse potential performance issues. Nevertheless, many systems still suffer from performance problems after they are deployed which are caused by the ever changing execution situation and loads in in-production environments. At this stage, there is a need  for advanced diagnostic tools that can detect performance bottlenecks by analyzing the behavior of the task execution along with the underlying deployed system.


Existing studies on root-cause analysis of performance anomalies can be grouped into two categories depending on whether they examine on-CPU or off-CPU \cite{perfcompass}. On-CPU analysis is a method that examines threads that are executing on the CPU. Therefore, this method is for measuring software performance and detecting bottlenecks caused only in the running-on-CPU duration. This, however, does not cover 100\% of the duration of the program. In the same way, typical debugging approaches such as cyclic profilers \cite{5740895} and interactive debuggers \cite{lehmann2018feedback} tend to be inefficient because thread schedulings and thread collaborations are outside of a profiler's scope, and debuggers require to pause the application under investigation.

In many cases, most of the runtime duration lies not inside the CPU, but rather in waiting for hardware resources and external factors such as the disk, I\/O, network, timer, or waiting for other threads to complete a task to release a lock. And if the time spent by threads when they are not running on CPU is synchronously used during an application execution, then it can impose latency and create performance issues \cite{gregg2013systems}. This is where off-CPU analysis, the focus of this paper, comes in to gain insight into thread scheduling, interactions among threads in different cores, and possible resource contention with the objective to identify the causes of latencies in multi-core systems \cite{wperf}.

In this paper, we introduce DepGraph, a novel approach for localizing performance issues in multi-core applications. Our method generates \textit{Waiting Dependency Graphs} from system-level traces with the objective to model all waiting relations that a thread may have during its critical path of execution. DepGraph is suitable to detect a variety of off-CPU performance issues including long-running system calls, blocking disk accesses, network latency, CPU scheduling issues, and causality collaborations among threads.

In addition, DepGraph relies on LTTng \cite{desnoyers2008lttng}, a powerful low-overhead Linux kernel tracing tool-set, making it scalable to large systems and production servers. We show the effectiveness of DepGraph by applying it to three different use cases where we were able to detect the root cause of different performance problems. 

The remainder of the paper is organized as follows: Section~\ref{related_work} discusses related works. Section~\ref{proposed_method} introduces the method architecture and the algorithm to build the dependency graph. Section~\ref{use_cases} shows the detection of latency causes in three real-world use cases. Section~\ref{evaluation} puts under the microscope the efficiency of the approach. Finally, Section~\ref{conclusion} concludes the paper.

\section{Related Work}
\label{related_work}

Execution tracing at the operating system level has been used for a variety of software and system performance analysis including automatic correlation of traces~\cite{wahab-correlation}, detection of code modification~\cite{abder1}, detection of erroneous configurations~\cite{Yuan2011,fdoray}, detection of bottlenecks at the disk layer~\cite{daoud2018recovering}, and program execution comprehension~\cite{systematicabstractionsurvey}. The analysis of kernel traces has been studied in a survey by Ezzati et al.~\cite{ezzatisurvey}. The book by Gregg~\cite{gregg2013systems} details a comprehensive method for detecting and diagnosing issues from kernel data. Notably, the author explains how to make a connection between the saturation of various resources with common problems.

Software performance analysis has been widely studied in the literature. One approach to evaluating software performance is to focus on metrics such as the number of external interrupts~\cite{julienD}, the number of scheduling per second~\cite{giraldeau}, the number of processes waiting for disk blocks~\cite{daoud2018recovering}, and the critical section pressure~\cite{davidCriticalPath} to identify the factors that affect performance. Giraldeau et al.~\cite{giraldeau} describe how runtime metrics about the processor, memory, file systems, and network usage, collected from Linux kernel traces allow diagnosing latency.

Other works approach latency issues directly and only analyse the trace spans in which the problem possibly occurred~\cite{giraldeau2016wait}. Significant on-CPU latency in a thread spending most of his running time on the CPU typically indicates a bad code design or an algorithmic problem. Such an issue is usually investigated and resolved using a \textit{profiler}. Profilers extract a sorted list of functions or modules which take the most execution time~\cite{gogleprofiler, syncprof}. However, profilers usually offer limited visibility and mostly work on user-space functions and modules. They may not be helpful for instances where the slowness comes from the execution of a system call. There are, however, system-level profilers with direct support from operating system kernel. For instance, Oprofile~\cite{oprofile} is a kernel-wide statistical profiling tool which regularly samples the program execution using a system timer. Oprofile is, however, not able to detect latency and to pinpoint its cause when the slowness is the result of interactions between multiple threads and resources.

Significant off-CPU latency such as waiting to obtain the CPU, to read a file, or to receive a network packet, mostly indicates a system overuse or a hardware-related problem. In that case, profilers cannot help identify the possible causes. Indeed, they only rank functions or modules based on statistical metrics without mentioning the nature of those issues~\cite{fournier2010analyzing}. Some researchers have considered the critical path analysis to get insight into off-CPU latency problems \cite{fournier2010analyzing, giraldeau2016wait}.

Critical path analysis was originally used in the field of project management to identify the longest tasks and to optimize them~\cite{willis1985critical}. Likewise, critical path analysis has been used in software debugging to extract bottlenecks in execution flows~\cite{fournier2010analyzing}. Fournier et al.~\cite{fournier2010analyzing} demonstrated the use of kernel events to understand the elapsed time details in a complex application. They rely on the \texttt{sched\_wakeup} event to identify the waiting states of user tasks for hardware resources. 

Giraldeau et al.~\cite{giraldeau2016wait} extended the method presented in~\cite{fournier2010analyzing} to support tasks distributed across multiple machines. Their work uses TCP packets matching to identify dependencies between such tasks. The critical path algorithm~\cite{giraldeau2016wait} estimates the minimal active path of a task and extracts different execution states along that path. Using their tool, one can see the state of a thread at different points in time and identify whether it is running or waiting for a resource. Their method does not show automatically which thread is currently using the blocking resource at the time of the waiting. Therefore, it still requires a manual investigation of the generated critical path to find the root cause. It shows that there is a possible blocking state but does not show why this is happening and which thread is responsible. Our proposed method is one solution to that problem. Indeed, DepGraph aggregates the critical paths of interacting threads, and by merging them into a history of resource usages, extracts the thread which is causing the blocking state. DepGraph allows comparing the different executions of the same task to directly pinpoint the cause behind similar latency issues.

Doray et al.~\cite {fdoray} also deployed the critical path algorithm~\cite{fournier2010analyzing, giraldeau2016wait} and extended it by adding stack traces collected from the userspace. Stack traces are gathered by unwinding the considered process, a common technique used by other profilers such as the Google PERF tool~\cite{gogleprofiler}. They populate a data structure called ECCT (Enhanced Calling Context Trees) from the stack traces collected along with the critical path segments and use that ECCT to compare the different executions of a single task. Results are visualized in an extended differential flame graph~\cite{gregg2013blazing}. 

Similar to ~\cite {fdoray}, other works have used user-level calling context combined with kernel traces to detect and debug latency problems. IntroPerf~\cite{introperf}, a Windows-based multi-purpose performance profiler, infers the performance of application functions in the call stack trace captured by the Windows kernel tracer (i.e., ETW). Although these methods are powerful, profiling and collecting stack traces can be costly and even impossible when the code is not accessible or is not compiled with debugging flags.

Although critical path analysis gives a precise breakdown of the elapsed time in general, it does not provide the reasons for the waiting states in the critical path. One would also like to know if waiting states are normal -- common to all similar executions. Our proposed method offers a solution to this problem. It provides all dependencies between the relevant threads and resources, regardless of whether they are within the critical path.

A recent work called wperf~\cite{wperf} also generates a wait-for-graph, to distinguish the local impact from the global impact of waiting events. They aim to identify events having large impacts on all threads, hence critical for total throughput. Our focus is however different. We focus on the performance of a single thread, along with all other system resources and threads in collaboration, in order to analyse and identify the reasons behind its unexpected latencies.

\section{The DepGraph Approach}
\label{proposed_method}

\begin{figure*}[ht]
\centering
\includegraphics[width=\textwidth]{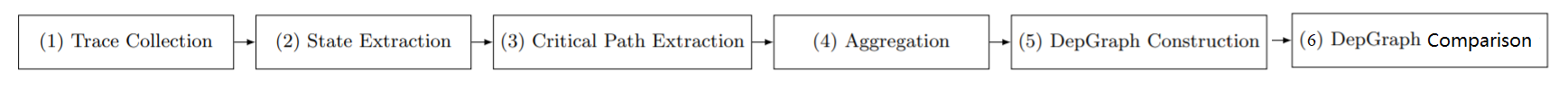}
 \caption{DepGraph construction steps.}
 \label{fig:pipeline}
\end{figure*}

The core of DepGraph is the extraction of all dependencies required to complete a task. Figure~\ref{fig:pipeline} shows the six steps of the DepGraph construction. 

First, we collect system-call traces from the kernel which characterize the thread execution. Because traces are known to be large, we need a way to reduce their size while keeping the essence of their content. To this end, we transform the trace into states. We consider a state to be the different status of a thread, such as running, waiting for disk, waiting for another thread, waiting for cpu, contented on a lock. The next step is to build a Waiting Dependency Graph (DepGraph) using state information to capture and display the different kind of the dependencies between threads. The last step is to group, merge, compare, and analyse the DepGraphs automatically to identify the root causes of latency problems. The steps are explained further in the subsequent sections.

\subsection{Trace Collection}

We use the \textit{Linux Tracing Toolkit: next generation} (LTTng) \cite{desnoyers2008lttng} tracer to collect system-level traces. LTTng is a well-maintained open-source tracing tool for Linux that allows tracing different layers including the kernel, the userspace, and virtual machines. LTTng is designed to have minimal overhead, which makes it suitable for performance analysis and in-production systems.

A trace collected using LTTng is a sequence of timestamped events representing an interaction between a set of system resources. For instance, the event \texttt{(tid1, syscall\_open, file1, cpu1, t1)} corresponds to a file open system call running on the CPU \texttt{cpu1} which is called by the thread \texttt{tid1} at time \texttt{t1}.

A trace usually encompasses the execution of several threads running concurrently in the system. When investigating performance problems, developers usually want to separately analyse either multiple tasks from a single thread (e.g., the chromium browser), a single task (e.g., opening a tab in the chromium browser), or a specific span of execution (e.g., compilation time of a Java application). Therefore, while debugging performance problems, one needs to specify the events that delimit the task(s) or behaviour of interest -- likely to contain the cause of the latency. For example, one could choose the kernel events \texttt{socket\_accept} and \texttt{socket\_shutdown} from the Apache thread to analyse a web request. Such delimiters define the start and the end of a request, respectively.

\subsection{State Extraction}

After collecting the trace data based on the task or behaviour of interest, we need to specify the events required to detect the waiting dependencies. To achieve this, we start by distinguishing the different states of the task to extract the ones related to waiting. As shown in Figure~\ref{exec_states},
a thread can be in one of four distinct states: ``running'' , ``runnable'' (preempted), ``blocked'', and ``interrupted''.

``Running'' denotes a thread executing code on a CPU. ``Runnable'' corresponds to a thread ready but unable to run due to a lack of CPU slices to assign. This happens when a higher priority thread is running on the CPU or when the current thread is out of its time slice. The state of a thread is ``interrupted'' when an IRQ or a softIRQ has occurred in its context and an interrupt handling routing is currently being executed on that CPU. Finally, a thread is ``blocked'' when it has been scheduled out and is waiting to acquire a resource or for some conditions to become true.

Blocking states are distinguished by their waiting cause, be it a system resource or another thread. Blocking states should not be readily mistaken for anomalies as they may be inevitable such as waiting for a request or a user input. In this paper, we study all blocking states causing delays to the thread execution.
\begin{center}
\begin{minipage}{.40\textwidth}
\begin{figure}[H]
 \centering
  \includegraphics[width=\linewidth]{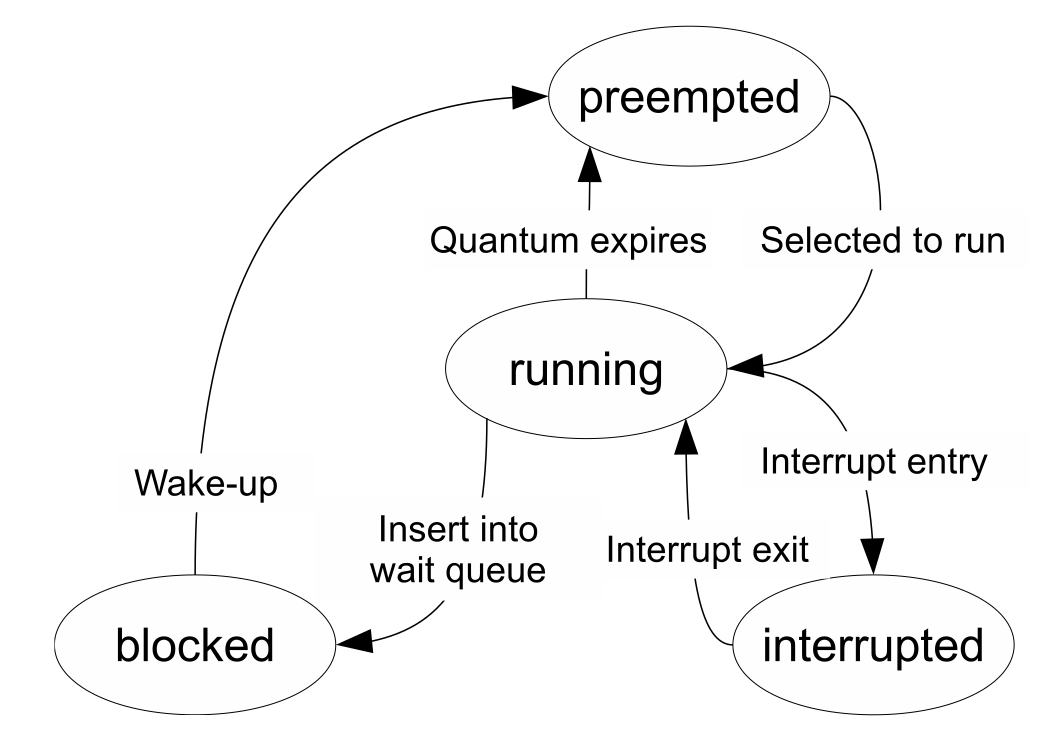}
  \caption{Execution states inferred from kernel events.}
  \label{exec_states}
\end{figure}
\end{minipage}\hfill
\end{center}
\begin{center}
\begin{minipage}{.40\textwidth}
\begin{figure}[H]
 \centering
  \includegraphics[width=\linewidth]{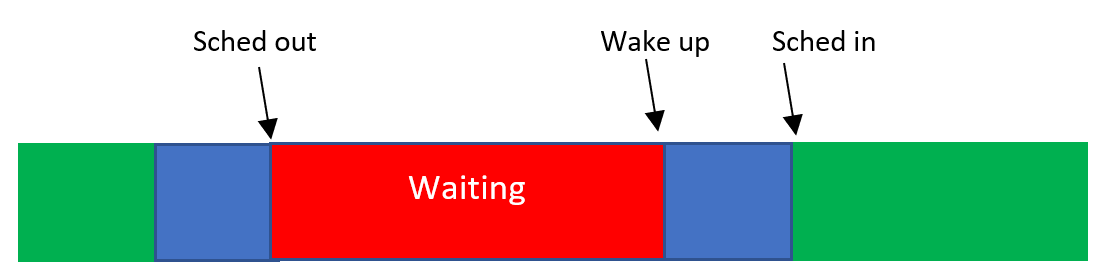}
  \caption{Waiting state extracted from raw kernel events.}
  \label{blocking_states}
\end{figure}
\end{minipage}
\end{center}
\bigbreak
To extract all thread states, we need the following system calls \cite {couturereport}:
\begin{itemize}
  \item \texttt{sched\_switch} indicates that a running thread is replaced by a new thread which is about to start running on the same CPU. This event produces two state changes in the system: it changes the state of the new thread to ``running'' and the state of the old thread to either ``runnable'' or ``blocked''.
  \item three entry/exit events indicate when a thread enters the interrupted state: \texttt{irq\_handler}, \texttt{softirq}, and \texttt{hrtimer\_expire} corresponding to the hardware, software, and timer interrupts, respectively.
  \item \texttt{sched\_wakeup} indicates that the state of a thread changes from ``blocked'' to ``runnable'', and its context reveals the blocking reason. For instance, calling this event from within a hardware timer context implies that a timer was blocking the thread. After being woken, the thread will wait to be chosen by the scheduler to run (see Figure~\ref{blocking_states}). 
\end{itemize}

Since a thread has various states during its execution phases, state values and the current system call at that time must be stored in a \textit{state database}. Otherwise, every time a state or system call is needed throughout the analysis, the trace should ineffectively be reread and reanalysed entirely.

State changes of other system resources are extracted with a similar approach. For instance, to track which thread is running on a CPU, a ``current thread on a CPU'' state value is created whenever a \texttt{sched\_schedule} event occurs on a CPU.

A \textit{state value} is an abstract event built from analysing some raw trace events. Formally, we define a state value $sv$ to be an event with a duration $d$ from timestamps $t_i$ to $t_j$, a key $k$ and a value $v$ associated with that key for that duration. State values are stored in a state database $SD$.

\begin{align}
&D = \{[t_i, t_j]\ |\ t_i, t_j \in \mathbb{N} , t_i < t_j\}\\
&SV = \{(d, k, v)\ |\ d \in D, k \in \text{Attributes}, v \in \mathbb{N}\}\\
&SD = \{sv\ |\ sv \in SV\}
\end{align}

Note that the key describes partially or entirely a system resource. For example, the key \texttt{fd} corresponds to a file descriptor -- an aspect of a system resource. Other examples of state values include ``thread running on a CPU'', ``thread using the disk'', ``thread waiting for a network connection'', ``CPU index on which a thread is running'', and ``CPU utilization of each thread''.

Every state-value type has a set of mapping rules $RL$ specifying the conversion from a set of trace events $e$ to a set of state values $sv$. Rules range from simple ``if then else'' to complex transition patterns depending on the analysis, the input events, and their effect on the system attributes state.

\begin{align}
&RL = \{(e \to sv)\ |\ e \subset \text{Trace}, sv \in SV\}
\end{align}

\subsection{Waiting Dependency Graph (DepGraph) Construction}

We consider a dependency to be any interaction between a thread and other threads or system resources. However, we only focus on \textit{waiting dependencies} (i.e., blocking states), since these are the ones that may cause delays and latency issues. We extract these dependencies from the execution traces, and use them to construct a \textit{Waiting Dependency Graph} (DepGraph). Since this graph is built from data collected at runtime, it may differ from what developers expected from their source code.

More formally, the DepGraph is defined as follows: given a set of threads and resources $S$, and a transition relation $R\subseteq S \times S$ with $(a, b) \in R$ denoting ``b depends on a'' or equally ``b is waiting for a'', the DepGraph is a directed graph $G=(S, T)$ with $T \subseteq R$. DepGraph is also acyclic provided that there are no deadlocks (i.e., ``a is waiting for b'' and ``b is waiting for a'').

Two types of dependencies are extracted: direct and indirect. Direct dependencies result from the wake-up relationship between threads. When a thread directly wakes up another, we know that the latter was waiting for the former. Such is the case when a web server thread emits a database query and the operating system changes its state to ``blocked'' while the control transfers to a database thread to fulfil the query. The database thread then wakes the web server thread after completing the query. This wake-up dependency is said to be direct and corresponds to an edge in the DepGraph.

\begin{algorithm*}[!ht]
 \caption{Algorithm to construct the DepGraph for a given thread}\label{dg}
 \begin{algorithmic}[1]
  \Procedure{DGraph}{$thread_A, SD, ts_s, ts_e, DG$}
  \State Input: source thread $thread_A$, state database $SD$, start timestamp $ts_s$, end timestamp $ts_e$, dependency graph $DG$
  \State $ES$ $\gets$ query $SD$ to get execution states of $thread_A$ in time range $(ts_s, ts_e)$
  \ForEach{$s \in ES $}
    \If{$state$ = ``blocked for another thread''}
     \State $curstate$ $\gets$ query $SD$ for current context of the $thread\_A$ (system call name, userspace or etc.)
     \If{$curstate$ is not null}
     \State $edg$ $\gets$ createEdge$(thread_A, \text{curstate}, ts_{s_{state}}, ts_{e_{state}})$
       \State $DG$ $\gets$ addToGraph$(edg, DG)$
     \EndIf
     \State $thread_B$ $\gets$ query $SD$ for other thread information
     \State $edg$ $\gets$ createEdge$(curstate, thread_B, ts_{s_{state}}, ts_{e_{state}})$
      \State $DG$ $\gets$ addToGraph$(edg, DG)$
      \State $DGT$ $\gets$ DGraph($thread_B, SD, ts_s, ts_e, DGT)$
      \State $DG$ $\gets$ mergeGraphs$(DGT, DG)$
    \ElsIf {$state$ = ``blocked for disk''}
      \State $curstate$ $\gets$ query $SD$ for current context of the $thread\_A$
     \State $T$ $\gets$ query $SD$ to get threads using the disk in time range $(ts_{s_{state}}, ts_{e_{state}})$
     \If{$curstate$ is not null}
       \State $edg$ $\gets$ createEdge$(thread_A, \text{curstate}, ts_{s_{state}}, ts_{e_{state}})$
       \State $DG$ $\gets$ addToGraph$(edg, DG)$
     \EndIf
     \State $edg$ $\gets$ createEdge$(curstate?curstate:thread_A, \text{``DISK''}, ts_{s_{state}}, ts_{e_{state}})$
      \State $DG$ $\gets$ addToGraph$(edg, DG)$
     \ForEach{$thread \in T $}
      \State $edg$ $\gets$ createEdge$(\text{``DISK''}, thread, ts_{s_{thread_{state}}}, ts_{e_{thread_{state}}})$
      \State $DG$ $\gets$ addToGraph$(edg, DG)$
      \State $DGT$ $\gets$ DGraph($thread, SD, ts_{s_{thread_{state}}}, ts_{e_{thread_{state}}}, DGT)$
      \State $DG$ $\gets$ mergeGraphs$(DGT, DG)$
     \EndFor
    \ElsIf {$state$ = ``runnable''} \Comment{waiting for cpu}
     \State $cpu$ $\gets $ query $SD$ to find the last CPU on which $thread_A$ was running
     \State $T$ $\gets $ query $SD$ to get threads running in $cpu$ in time range $(ts_{s_{state}}, ts_{e_{state}})$
     \State $edg$ $\gets$ createEdge$(thread_A, \text{``CPU''}, ts_{s_{state}}, ts_{e_{state}})$
      \State $DG$ $\gets$ addToGraph$(edg, DG)$
     \ForEach{$th \in threads $}
      \State $edg$ $\gets$ createEdge$(\text{``CPU''}, th, ts_{s_{th_{state}}}, ts_{e_{th_{state}}})$
      \State $DG$ $\gets$ addToGraph$(edg, DG)$
      \State $DGT$ $\gets$ DGraph($thread, SD, ts_{s_{th_{state}}}, ts_{e_{th_{state}}}, DGT)$
      \State $DG$ $\gets$ mergeGraphs$(DGT, DG)$
     \EndFor
    \EndIf
  \EndFor
  \State \textbf{return} $DG$\Comment{The dependency graph}
  \EndProcedure
 \end{algorithmic}
\end{algorithm*}

\begin{algorithm}[!htp]
 \caption{To add a new edge to an existing DepGraph}\label{merge}
 \begin{algorithmic}[1]
  \Procedure{addToGraph}{$edg, DG$}
  \State Input: edge $edg$ , dependency graph $DG$
  \If{$edg \in DG$}
    \State $newlabel_{edg} \gets label_{edg} + label_{edg\_\text{in}\_DG}$
    \State $DG$.UpdateLabel($edg, newlabel_{edg}, DG$)
  \Else
    \State $DG$.InsertEdge($edg, DG$)
  \EndIf  
  \State \textbf{return} $DG$\Comment{The dependency graph}
  \EndProcedure
 \end{algorithmic}
\end{algorithm}

Indirect dependencies occur when threads are competing for a resource without direct wake-up call between them. For instance, an indirect edge is created when a thread is waiting for a CPU used by another thread. The state database allows finding the thread which is using the resource; thus the thread responsible for the waiting. in this study, only two types of competition are considered: ``waiting for CPU'' and ``waiting for disk''.

Upon entering a blocking state, the current system call is extracted from a state history database (line 6 in Algorithm~\ref{dg}). Provided that the waiting happens within a system call -- between the start and end events delimiting its execution -- the method creates a node in the graph for that system call.

Whenever an edge is to be inserted in the graph, the method \texttt{addToGraph} checks if that edge already exists. If it exists, its weight is updated by summing the previous value with the new one (lines 4 and 5 in Algorithm~\ref{merge}) creating an aggregation of the labels and times.

Whenever there is a direct or indirect dependency between two threads, the DepGraph of the running thread is created for the duration of the dependency. The graph is then merged with the one of the waiting thread (lines 15, 29, and 40 in Algorithm~\ref{dg}). The function \texttt{mergeGraphs} implementation is similar to the one of \texttt{addToGraph} (Algorithm~\ref{merge}) and has been omitted for the sake of brevity.

\section{Analysis of the DepGraph}



The DepGraph, in general, exposes an aggregation of the different waiting states at runtime. Comparing the DepGraphs of different execution groups, e.g., fast and slow executions, allows explicitly pointing out where the longest waiting periods occur and where the difference starts. Such comparison of DepGraphs is particularly useful to identify the reason behind slow executions, an important question for various applications. 
Let us now demonstrate the DepGraphs construction in an example in order to illustrate how a DepGraph is used in a real performance evaluation scenario. Grouping and comparison of DepGraphs will also be explained shortly.



\begin{figure*}[!ht]
\centering
        \begin{subfigure}[b]{0.37\textwidth}
            \centering
            \includegraphics[width=\textwidth]{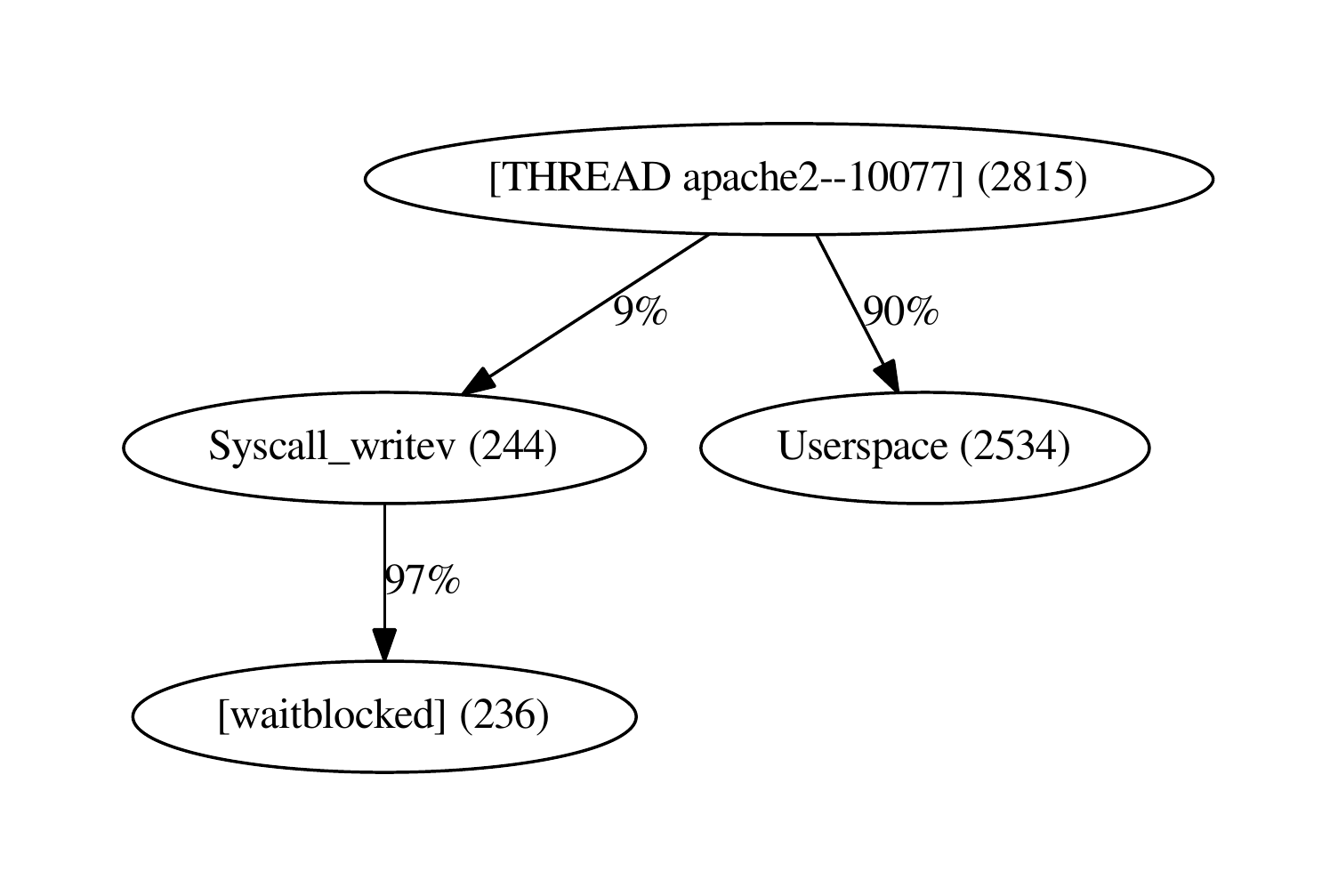}
        \end{subfigure}
        \begin{subfigure}[b]{0.40\textwidth}  
            \centering 
            \includegraphics[width=\textwidth]{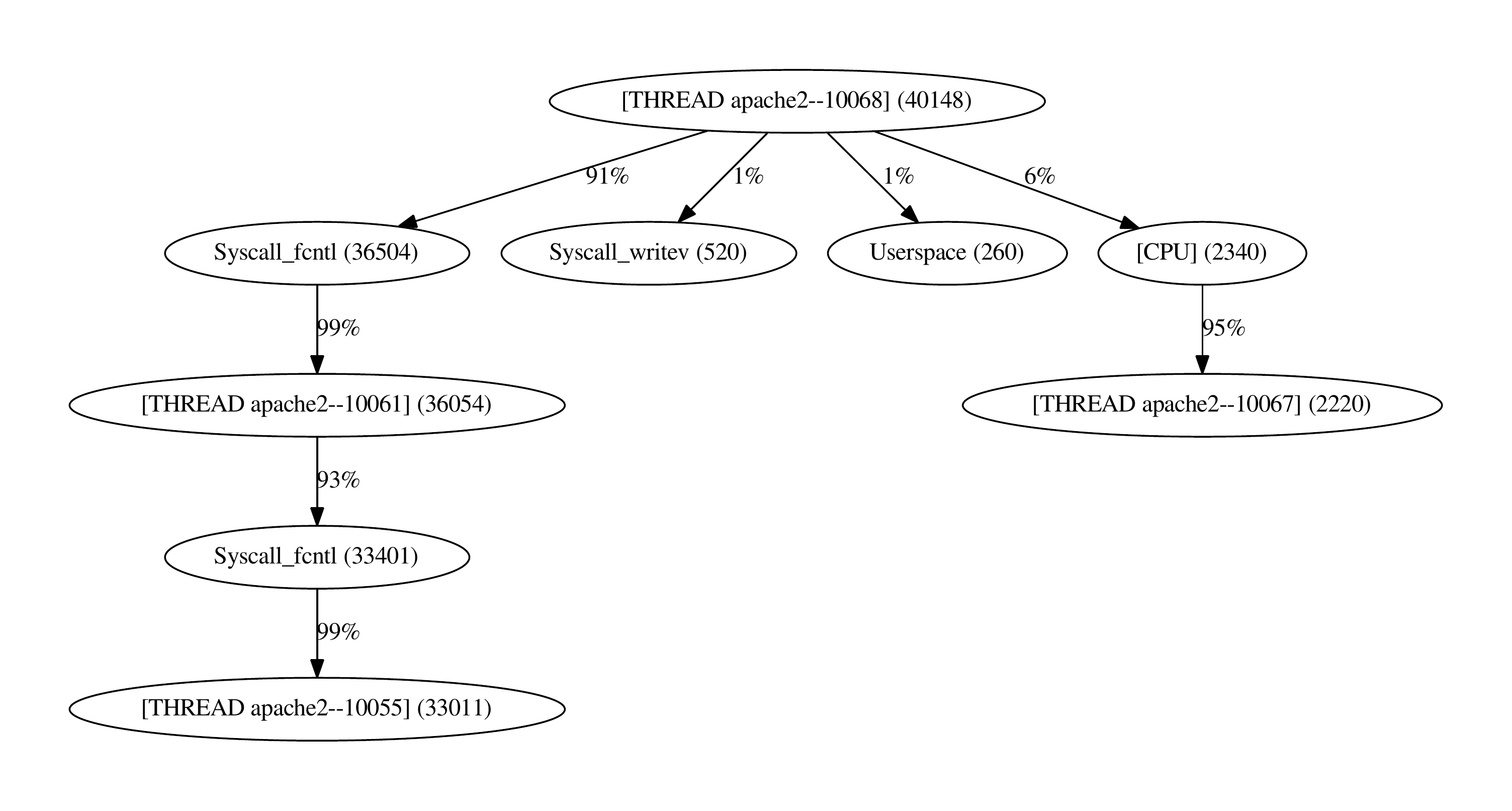}
        \end{subfigure}
    \caption{The DepGraphs showing the lock contention between Apache processes occurred within fcntl system call context.}%
    \label{lock_contention}
\end{figure*}

Figure~\ref{lock_contention} shows the DepGraph of a ``fast'' web request on the left and a ``slow'' web request on the right. Using the DepGraph, we want to figure out what is the reason behind the slowness of the second request. In the DepGraph, the root node always corresponds to the running thread for which we are undertaking the root-cause analysis. In this example, the execution thread is Apache2 with a thread number 10077 for the fast request, and 10068 for the slow one. Note that the thread number is not particularly useful in interpreting the DepGraph in this case. To the right of the thread name, in between the parentheses, is the total execution time in microseconds. As expected, the slow request has an execution time more than ten times greater than the fast one (40148 $\mu s$ vs 2815 $\mu s$).
Moving down the graph, we notice that the fast request spends 90\% of the time in the userspace, which only takes 2534 $\mu s$ to execute. Compare this to the slow request which spends 91\% of its execution in a \texttt{syscall\_fcntl}\footnote{\texttt{syscall\_fcntl} allows controlling a file, most notably placing a read or write lock.}, and the cause of the latency is clear: there is a lock contention. This use case is presented in greater detail in Section~\ref{sec:lock_contention}.

\subsection{Grouping, Merging and Comparing the DepGraphs}
So far we have generated a DepGraph for each execution, now, the goal is to use these graphs to find the root cause of slow performance. First, we group the executions based on
parameters that we extract during the trace analysis. It is important to note that these
parameters represent different runtime aspects of a given execution, such as the execution time, the number of page faults, number of I/O, number of disk
accesses, and number of interrupts. This also applies to the time spent waiting for a
specific event such as waiting for CPU, waiting for disk access, waiting for a packet to be
received or sent, or waiting for a timer. To group the execution we use k-means clustering which is a useful and simple tool for grouping the executions based upon our chosen
features.

For each execution, we extract some runtime data and use them as input values for the k-means algorithm. This allows us to accurately cluster executions represented by DepGraphs. If a majority of our graphs depict executions that run normally then the k-means clustering algorithm can easily detect outliers with abnormal values for our chosen feature.

In this method, for each execution the count and duration of different execution states like Blocked for disk I/O, Blocked for CPU, Blocked for futex, Blocked for a task, Blocked for interrupt, Blocked for timer, and also parameters like number of page faults, number of bytes read/written, total request time are extracted and used as the feature sets for the k-means clustering algorithm in order to group the extracted DepGraphs of executions. The number of clusters is configurable but we use 2 to 4 in our experiments to get  various groups of executions to be used in the comparison phase. There are heuristics to compute the ideal number of clusters, however not addressed herein and is left for a future work.

The root-causes of performance issues can be found from comparing the DepGraphs of the different clusters, e.g., normal and abnormal executions. To do so, we first create a representative model for each cluster and then we pass them to a comparison algorithm. Representative model comes from merging the graphs within a cluster into a single graph while simultaneously merging their nodes. 

In our work we chose to compare only two clusters at a time as our goal is to derive the difference between two specific groups of data. For instance, we may want to look at the
difference between executions that waste CPU cycles or have a poor response time in contrast to normal executions. Therefore, comparing more than two groups at once is unnecessary as the
information we are seeking can be found in the two clusters chosen.

\begin{figure}[H]
 \centering
  \includegraphics[width=\linewidth]{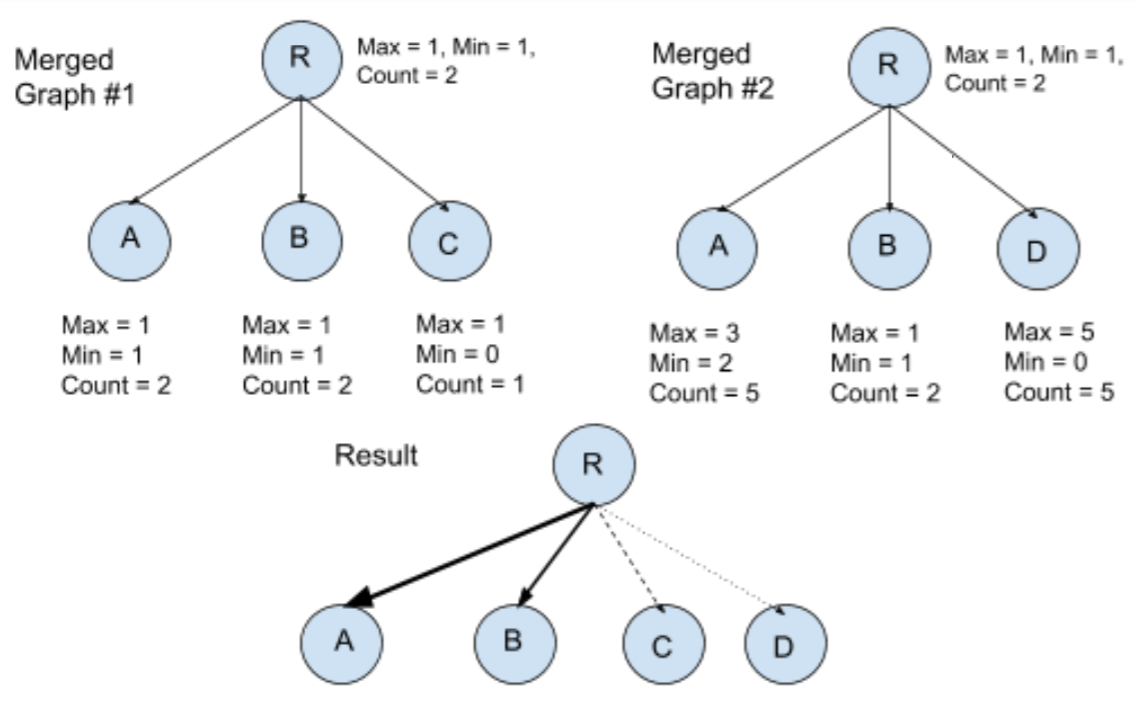}
  \caption{Comparing the two representative DepGraphs.}
  \label{comparinggraphs}
\end{figure}

Figure \ref{comparinggraphs} shows a visual representation of the process of comparing two representative graphs that were generated from two different clusters. To compare two graphs, the left and right mean’s are calculated based on their count values. Next, the standard deviation for each left node is calculated. Finally the Mean Difference's Standard Deviation is used to determine a boldness level for the edges.

We have defined five different levels of boldness that are determined from the level of difference between graph 1 and 2. The boldness level chosen depends on just how different the second graph is from the first meaning we can represent various levels of difference between two graphs and their nodes.

The resulting graph uses dashed lines to show calls (edges) that only occurred within the first graph, dotted lines for calls (edges) that only occurred within the second graph, and normal lines for calls that occurred in both. These normal lines have boldness levels according to the difference in node values. For example, in Figure \ref{comparinggraphs} the resulting graph’s node A has a much bolder edge as the count increases suddenly from graph 1 to 2.

The resulting graph, shows the main differences between two different groups which can reveal the root-cause of the problem. For example, comparing the normal web requests to the very slow web requests can give us a graph having nodes/edges (e.g., waiting for the CPU or waiting for a timer, etc.) which appear only in the slow requests, or a graph showing the differences between different execution states, like waiting more/less for CPU, for a timer, for disk driver to complete a task, for another task (e.g., database thread) to complete a query, etc. The graph resulting from the comparison of the two groups can then highlight the main difference(s) between the groups,  revealing the actual root-cause(s) behind the problem. We will see some examples in the next section. 
\section{Use Cases}
\label{use_cases}

This section demonstrate how analysing and comparing the DepGraphs allows highlighting the root cause of performance issues in three distinct use-cases.

\subsection{Lock Contention}
\label{sec:lock_contention}

A web application written in PHP is running on a server comprising of Apache 2.4.23 web server, MariaDB 10.2 database server, and PHP 7.0 installed as an Apache module. Periodically, the request response time increases. A careful inspection of the source code provides no possible cause as the program runs with an acceptable performance most of the time. The database design seems good since single queries run within the expected time range.

The comparison of DepGraphs shows clearly that in the slow requests there is an excess of interactions between some Apache threads which are absent from the DepGraph of normal requests (see Figure~\ref{lock_contention}). This observation indicates that Apache threads are suffering from some unexpected contentions. As shown in Figure~\ref{lock_contention}-b, the contention occurred within the \texttt{fcntl}\footnote{The \texttt{fcntl} system call manipulates a file descriptor, see \url{http://man7.org/linux/man-pages/man2/fcntl.2.html}} system call, where $82\%$ of the waiting took place. This contention is unexpected considering that the Apache workers are supposed to run PHP scripts independently. A deeper investigation of this inter-thread collaboration highlights a cache issue. Indeed, PHP employs at its core a performance-enhancing cache mechanism called OPcache (OpCode Cache) to cache compiled script byte-codes in shared memory. Whenever a script needs to be compiled, the thread checks the OPcache on the shared memory for an already-compiled code. If the code is missing, the PHP engine compiles the code and writes the resulting byte-code in memory. 

Assuming that the OPcache is enabled (which is the default setting in PHP 7.x) and that the cache is empty or freshly restarted, the PHP engine tries to compile, optimize, and write OpCode in the shared memory for each concurrent request. This operation requires to exclusively lock the whole shared memory. This dependency appears in the graph because PHP runs within the Apache threads -- remember it was installed as an Apache module. Once a thread takes the lock, every other threads have to wait before they can obtain the lock and write their own compiled OpCode in memory. This explains the waiting dependencies between threads shown in Figure~\ref{lock_contention}.

By further looking at the PHP configuration, we figured out that the size of the OpCode Cache (OPcache) is the limiting factor. Increasing the cache size decreases the number of cache restarts and therefore solved the problem. Indeed, the new cache size was greater than the memory required to store all compiled script byte-codes. Therefore it removed the periodic cache restarts, hence the need for periodic compilations causing the lock contention and latency. Using the duration-based features to cluster the requests into fast and slow, constructing the DepGraphs of different groups and comparing them reveal an unexpected collaboration in single group of requests (the slow requests). Therefore, the source of the latency is a hidden lock contention between Apache threads.

\subsection{CPU Contention}
A real-time task called \texttt{ktimersoftd} which is scheduled to run many times with \textit{Cyclictest}\footnotetext{\url{https://wiki.linuxfoundation.org/realtime/documentation/howto/tools/cyclictest/start}} on a real-time Linux kernel takes longer (9 ms) than usual (1 ms) to complete in certain instances.

Without any prior knowledge about the latency root cause, we apply our method to gain insights into the problem.

\begin{figure}[!ht]
    \centering
      \includegraphics[width=0.37\textwidth]{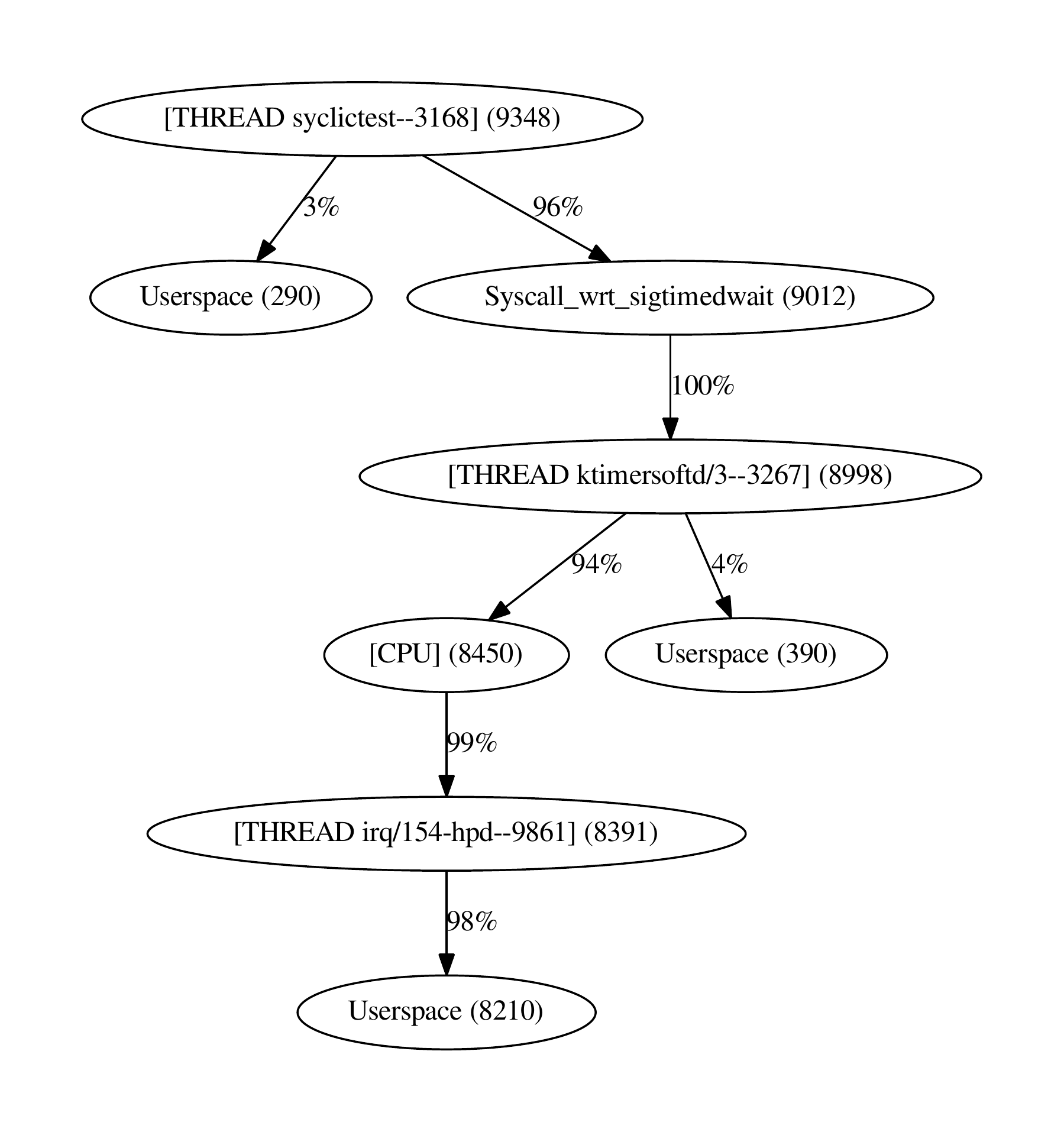}
    \caption{The DepGraph of a real-time task revealing a CPU contention problem.}
    \label{cpu_contention}
\end{figure}

We fully trace the benchmark execution and we apply our method on the trace collected from the slow periods. Although the source code of the real-time task is open-source, we did not modify the code and instead rely solely on the trace data collected from the kernel while the benchmark was running.

After grouping the executions into two different normal and abnormal groups, collecting the trace data for each group, generating the DepGraphs (an example is shown in Figure~\ref{cpu_contention}) for each execution and comparing them reveal a blocking dependency between the real-time task \texttt{ktimersoftd/3--3267} and a higher priority task woken by an interrupt \texttt{irq/154-hpd} in the slow executions. This do not happen in normal executions where the benchmark is fully running on CPU without any interruptions and no significant CPU usage by other threads.

Although the Cyclictest has a higher priority than the interrupt handling thread, the same is not true for the real-time task. Instead of running immediately, the task has to wait to acquire the CPU which is busy with the interrupt handling thread. The interrupt \texttt{irq/154-hpd} is actually called when an HDMI cable is plugged to the machine.

Note that the Cyclictest tool easily detects and reports the unexpected latency. However, it is not able to show why this is happening and what (e.g., which thread) is responsible for this latency. Our method explicitly reveals that the main reason behind this latency is an issue with an interrupt handling thread (\texttt{irq/154-hpd}) with a higher priority than the real-time task which makes the CPU busy and prevents the real-time task from using the CPU on time.

This use case clearly shows that the proposed method is able to detect CPU contention problems without using the source code and is also able to provide a hint of the thread(s)/resources responsible for the contention problem.

\subsection{Disk Contention}

The performance of web requests handled by an Apache web server has been monitored. Although most of the requests are executed in an acceptable amount of time, some are processed about $10$ times slower. 

\begin{figure*}[!ht]
 \centering
   \includegraphics[width=.80\textwidth]{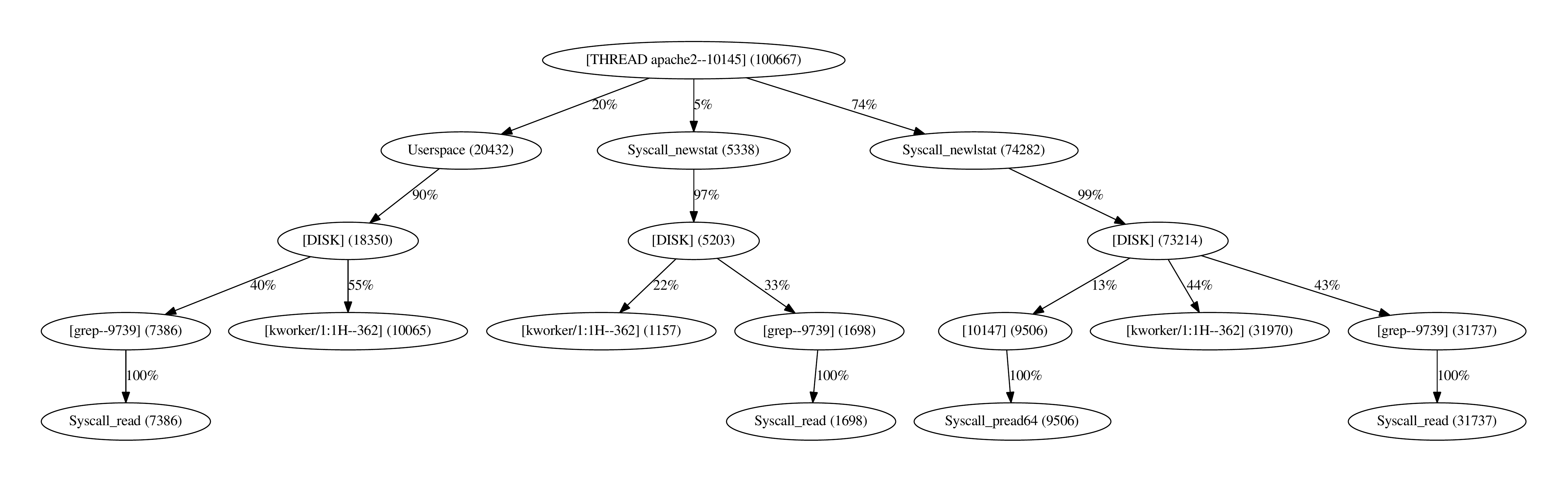}
   \caption{Disk contention caused by a recursive grep command.}
   \label{usecase_apache_disk}
 \end{figure*}

The DepGraphs comparison shows that in the slow requests the Apache threads spend most of their execution time waiting for file-related system calls. For instance, a slow request handling in the Apache thread \texttt{10145} spent most of its time running the \texttt{newstat} and \texttt{newlstat} system calls which are hindered by the disk being busy with other threads (see Figure~\ref{usecase_apache_disk} for an example).
x`
The \textit{grep} thread \texttt{9739} is using the disk for $43\%$ of the \texttt{newfstat} execution, causing the Apache thread to wait more than usual to fetch the required script from the disk. The \texttt{newfstat} system call is called every time a file is requested to verify that it actually exists and that it is a regular file (i.e., not a directory). The DepGraph also reveals that the disk is busy with two other threads (\texttt{kworker} and \texttt{10147}) while \texttt{newstat} is running.

This disk contention is clearly the main reason behind the observed web request slowness.

\section{Evaluation}
\label{evaluation}

In this section, we discuss the efficiency of the proposed method. We first detail the methodology to measure the tracing cost. We then evaluate the time required to extract the execution model and generate the Waiting Dependency Graph.

\subsection{Setup}

Traces were collected using LTTng 2.11 on a workstation equipped with 32 GB of RAM and a quad-core Intel® Core™ i7-6700K CPU @ 4.00 GHz. The operating system was Linux Ubuntu 16.04.6 LTS with the 64-bit kernel 4.15.0-62. The local hard disk reference 7200 RPM WDC WD1003FZEX-0 stored the trace data and trace analysis results.

\subsection{Tracing Cost}

To measure the tracing overhead, four different configurations were considered: 
\begin{itemize}
  \item Base case: tracing is disabled.
  \item Full tracing: all kernel tracepoints are enabled. This worst case is not required by our method. Results are however provided to show the cost of full Linux kernel tracing. 
  \item System call tracing: only system call events are traced. System call tracing has been and still is extensively used. The overhead of the data collection process is however rarely discussed.
  \item Dependency tracing: only system calls and events required to extract the dependency states and to generate the dependency graph are enabled. This is our method's setting.
\end{itemize}

We generated three benchmark profiles with the \texttt{sysbench} tool that represent most of the use cases for our method:
\begin{itemize}
  \item \texttt{sysbench} CPU test computes prime numbers up to $20000$. In this benchmark, the program runs mostly in userspace and does not include system calls. The only kernel events are related to CPU scheduling. 
  \item \texttt{sysbench} IO test calls a great number of system calls.
  \item Interleaving of \texttt{sysbench} CPU and IO tests to simulate multithreaded applications such as MySQL, MongoDB, Apache, etc.
\end{itemize}

Table~\ref{tracingcost} shows the overhead of the above benchmarking profiles for each tracing setting. Results are the average of $50$ executions.

\setlength{\tabcolsep}{8pt}
\begin{table}[!ht]
\centering
\caption{Relative duration indicating the tracing overhead of different profiles}
\begin{tabular}{lrrr}
Benchmark    & CPU  & IO    & Multithread \\ \hline
No tracing   & 100.0\%  & 100.0\%    & 100.0\% \\
Full tracing    & 101.0\%  & 142.3\%    & 123.5\% \\
Syscall tracing  & 100.5\%  & 108.7\%    & 107.3\% \\
Dependency tracing & 100.7\%  & 110.1\%     & 108.5\% \\ \hline
\end{tabular}
\label{tracingcost}
\end{table}

The results show that the overhead imposed by tracing while the system is busy with CPU computations never exceeded $1\%$ regardless of the tracing setting -- as measure by the benchmark duration.

The results are drastically different for the IO benchmark. A large number of mostly file-related system calls are generated. Full tracing imposed $42\%$ overhead due to the trace itself generating many disk IO operations, thus competing with the benchmark over the disk. To confirm this, the benchmark was repeated while the trace was sent over the network to a relay daemon running on a remote system. The amount of local IO operations was reduced and the worst-case scenario overhead dropped to a reasonable $6\%$.

The mixed benchmark behaved better than the IO one in terms of the tracing overhead. In that case, the number of system call events is reduced by $40\%$ and the overhead of the worse case was in the order of $23\%$. In all three benchmarking use cases, the overhead of the dependency tracing mode which is our method's setting never exceeded $10.1\%$.

\subsection{Analysis Cost}

The collected traces were read with \texttt{Babeltrace} \footnote{https://github.com/efficios/babeltrace}, a C library that decodes the CTF trace format. CTF is a common trace format that LTTng and many other tracers used to encode their tracing data. As expected, the analysis duration positively correlated with the trace size -- with the number of events rather than its duration. To build the DepGraph, the trace has to be read only once. Thus, analysis data including state changes and dependencies is generated and stored in the state history database, an efficient interval tree that handles data in the order of terabytes. Grouping and comparison times are dominated by the DepGraph construction time. Therefore, they are neglected in this analysis.

\begin{table}[!ht]
\centering
\small
\caption{Trace reading time and analysis cost.}
\begin{tabular}{lrrr}
Benchmark        & CPU & IO & Multithread \\ \hline
Tracing time (s)     & 25 & 12 & 15 \\
Trace Size  (MB)    & 31 & 104 & 74 \\
State Database Size (MB) & 4.6 & 9.1  & 8.4 \\ 
Trace Reading Time (s)  & 10.8 & 20.87 & 17.1 \\
Analysis Time (s)    & 4.2 & 7.1 & 6.5 \\\hline
\end{tabular}
\label{analysiscost}
\end{table}

The evaluation reveals that the analysis time is constantly between $30\%$ and $40\%$ of the time required to read the trace data. Since the analysis is done while reading the trace, the reading time dominates the analysis time. Table~\ref{analysiscost} details the analysis overhead only for the dependency tracing setting.

Table~\ref{analysiscost} also shows the state database size which was used to store and populate the internal states of the threads and system resources. As mentioned previously, our method uses the tree-based data structure proposed by~\cite{alexshtconference} to store the history of system states along the time. 

The state database works as a lookup table in our method so that the dependency graph generator is able to extract the states of the main thread, correlating threads, and resources. Since the database employs a disk-based tree data structure, it offers a fast query time (order of $O(log n)$) and can store a very large state history. Although our method is independent of the internal implementation of this lookup database, we evaluate the cost of using a spatial database like PostgreSQL to see if it is possible to get better performances. Note that PostgreSQL is a database system that allows managing multi-dimensional spatial data.

\begin{center}
\begin{figure}%
    \centering
    
    \subfloat[Disk usage]{{\includegraphics[width=0.31\textwidth]{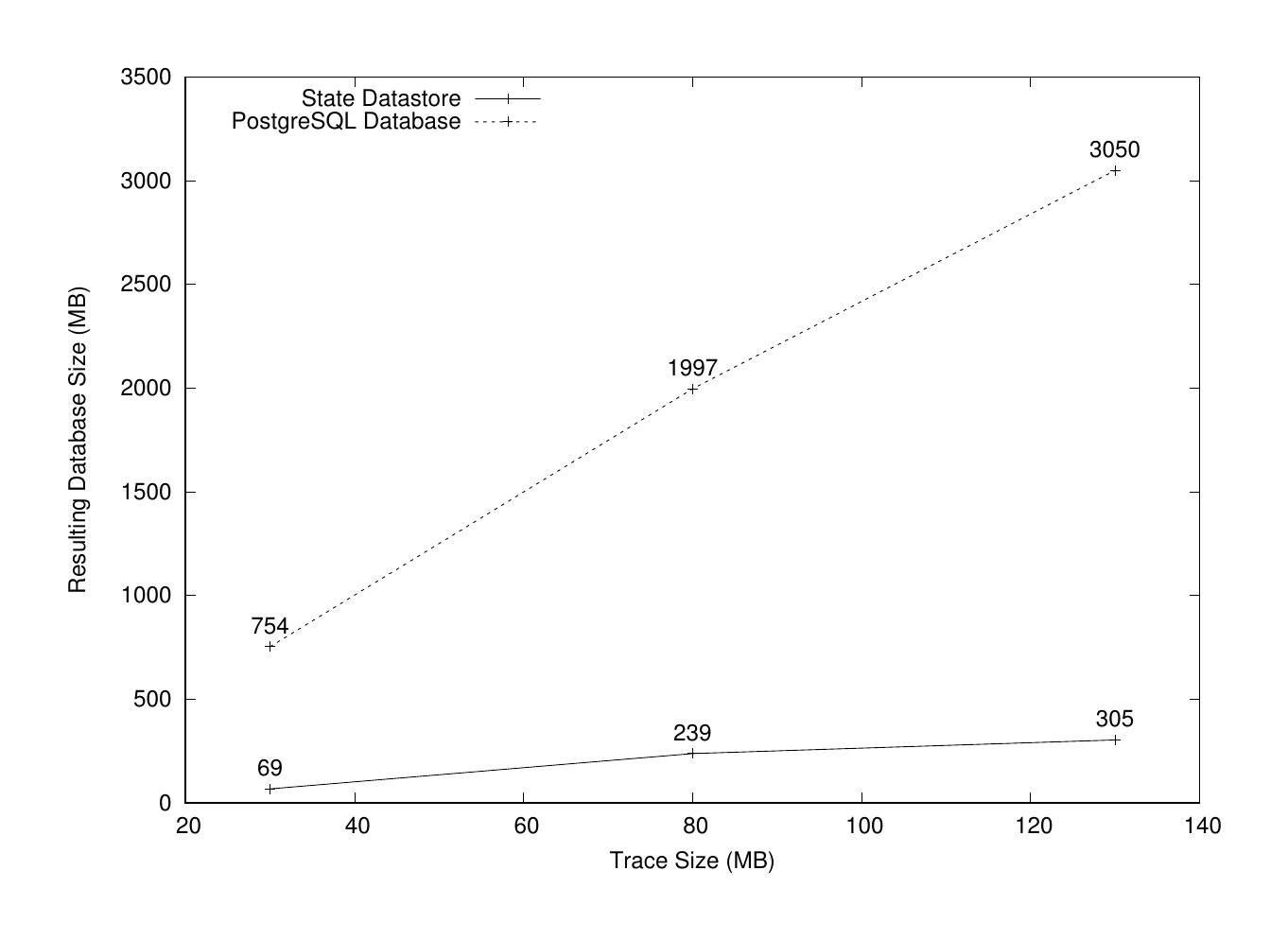}}}\\
    \subfloat[Processing time]{{\includegraphics[width=0.30\textwidth]{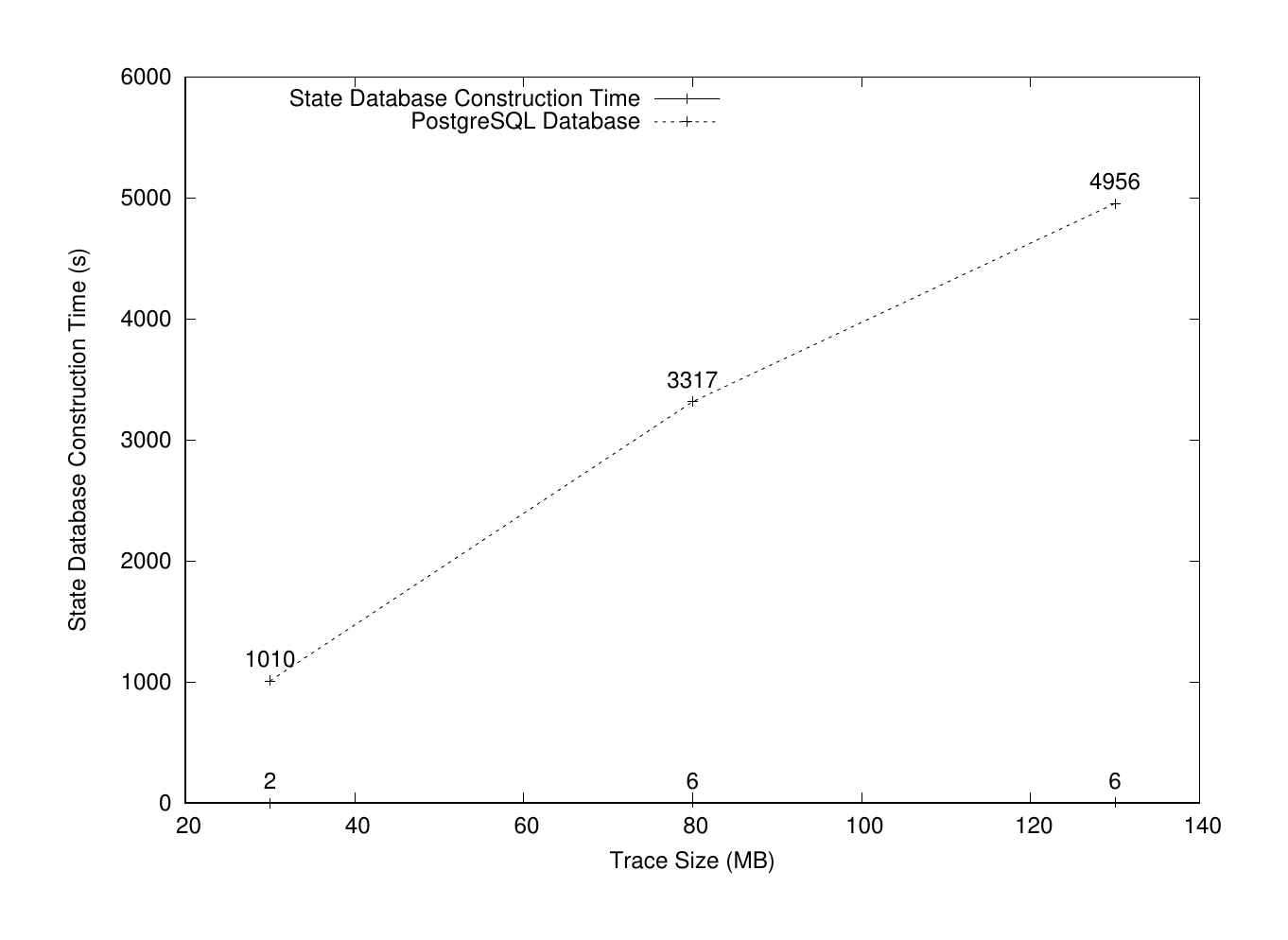}}}\\
    \subfloat[Query time]{{\includegraphics[width=0.31\textwidth]{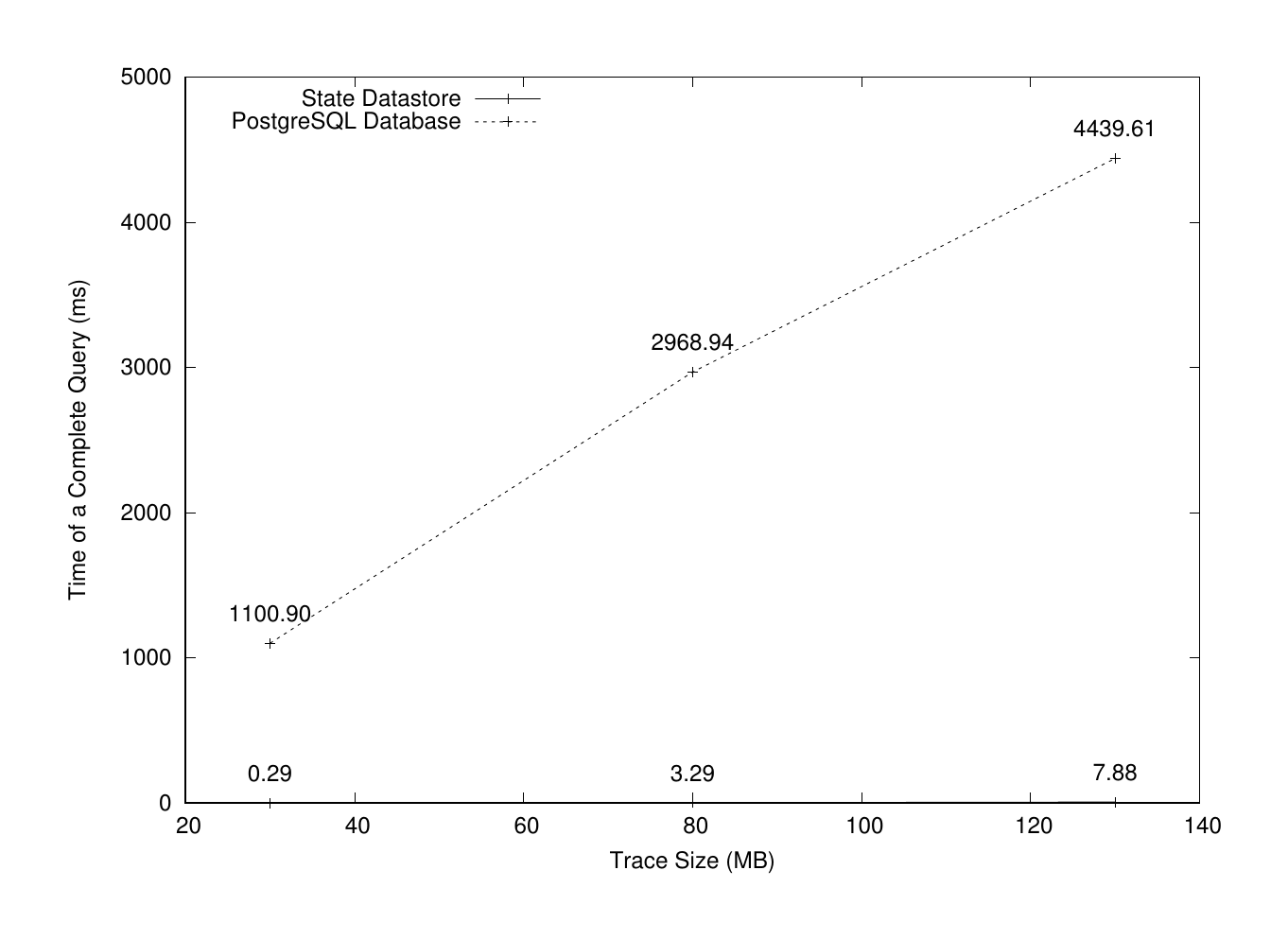}}}
    \caption{Comparison between the state database tree and PostgrSQL.}%
    \label{fig:StatePostgre}
\end{figure}
\end{center}
Figure~\ref{fig:StatePostgre} shows that PostgreSQL takes much longer to construct than the state database, especially with larger trace files. Furthermore, PostgreSQL requires more space to save the data and also takes longer to serve queries. Note that to test the disk usage, we must keep track of a large number of different states for all running threads and resources.

The results above demonstrate that even if PostgreSQL provides highly optimized mechanisms to manage generic data-sets, it cannot outperform a specific-purpose database such as the interval tree. The interval tree is tailored for trace files and stores the data in ways that make it quickly accessible while requiring minimal disk space.
 
\section{Conclusion}
\label{conclusion}

In this paper, we proposed a method to model the waiting dependencies between threads and hardware resources solely from kernel-level tracing events. Thus, the Waiting Dependency Graph (DepGraph) method is usable for performance analysis of a large set of applications. This graph allows developers to investigate execution behavior and interactions extracted at runtime, and compare them with the expected behavior. By doing so, the method highlights possible unforeseen bottlenecks and the reasons behind them which helps optimize the code or the system configuration, or improve resource management and efficiency in general.

We illustrated the usefulness and efficiency of the proposed method in different use-cases and configurations. The results show that the tracing and creating the DepGraph impose a total overhead lower than $10.1\%$, making the method suitable for performance anomaly and in-production environments.

\section{ACKNOWLEDGMENT}
We would like to gratefully acknowledge the Natural Sciences and Engineering Research Council of Canada (NSERC), Ericsson, Ciena, and EffciOS for funding this project.


\end{document}